\documentstyle[12pt]{article}
\begin{document}
\setlength{\baselineskip}{24pt}
\title{Color Magnetic Corrections to Quark Model Valence Distributions}
\author{C.J. Benesh, T. Goldman,\\
Theoretical Division, MSB283, \\
Los Alamos National Laboratory, \\
Los Alamos, NM 87545,\\
and,\\
G.J. Stephenson Jr.,\\
Physics Division, MSD434,\\
Los Alamos National Laboratory,\\
Los Alamos, NM 87545 }
\maketitle

\begin{abstract}
	We calculate order $\alpha_s$ color magnetic corrections to the
valence quark distributions of the proton using the Los Alamos Model Potential
wavefunctions. The spin-spin interaction breaks the model SU(4)
symmetry, providing a natural mechanism for the difference between the
up and down distributions. For a value of $\alpha_s$
sufficient to produce the $N-\Delta$ mass splitting, we find up and down
quark distributions in reasonable agreement with experiment.
\end{abstract}

\section{Introduction}
	Historically, quark models have provided us with a convenient,
if simplistic, method for making quantitative calculations of
low energy hadronic observables. Unfortunately, much of the
available data on hadronic structure is taken in an energy
regime well beyond that at which simple quark models are expected
to be valid. In particular, the distribution of quarks in the nucleon
as a function of their light cone momentum fraction has been measured
in numerous experiments over a wide energy range in the scaling regime.
In order to take advantage of this data, it was necessary to invent
an argument allowing an extrapolation from the low energy quark model regime
to the high energies where direct measurements can be performed.

Such an
argument was put forward by Jaffe and Ross\cite{1} in 1980. Quark models,
they argued, could only be a representation of QCD at a relatively low
renormalization scale, $\mu^2\approx 1/R_{proton}^2$, as a result of their
relative simplicity. More explicitly, they argued that, at large $\mu^2$,
experiments
are able to resolve structures of order $1/\mu$, and consequently one sees
a hadron as a very complicated object, composed of valence quarks, lots
of glue, and a sizeable sea. As the renormalization scale gets smaller,
the resolution decreases,
sea quarks get recombined into gluons, and gluons are reabsorbed
by quarks. Indeed, one can speculate that at very low renormalization scales,
virtually all of the sea and most of the glue can be reabsorbed into
the valence quarks, so that one is left with a proton composed of three
``constituent'' quarks bound together by some residual interaction, a quark
model\cite{1a}!
While this argument says nothing about which of the great variety of
quark models describes the nucleon, it provides a well-defined prescription
for testing models against high energy data. One simply calculates
quark distribution functions using the model and then uses the renormalization
group to evolve the result from the quark model scale, $\mu_0^2$, to a scale
relevant to experiment.

	The first step is to calculate the quark distribution at the quark
model scale. For unpolarized scattering, the spin averaged quark distributions
can be written, in the nucleon rest frame, as\cite{2}
\begin{eqnarray}
q_i(x)&=&{1\over 4\pi}\int d\xi^- e^{iq^+\xi^-}
<N\vert \bar\psi_i(\xi^-)\gamma^+{\bf P_+(\xi^-)}\psi_i(0)\vert N>\vert_{LC}
\nonumber \\
& &\nonumber\\
\bar q_i(x) &=& -{1\over 4\pi} \int  d\xi^- e^{iq^+\xi^-}
<N\vert \bar\psi_i(0)\gamma^+{\bf P_-(\xi^-)}\psi_i(\xi^-)\vert N>\vert_{LC},
\nonumber\\
& &
\end{eqnarray}
where $q^+=-Mx/\sqrt{2}$, with $x$ the Bjorken scaling variable,
$\psi_i(\bar\psi_i)$ are quark field operators of flavor $i$, $\gamma^+=
(\gamma^0+\gamma^3)/\sqrt{2}$ is a Dirac gamma matrix, $\bf P_\pm(\xi^-)$ is a
path ordered
exponential, $\exp[{\pm ig_s\int_0^{\xi^-} A^+(\eta^-) d\eta^-}]$ with $g_s$
the strong coupling constant, that insures the gauge invariance of the
operator, and the subscript
$LC$ denotes the light cone condition on $\xi$, namely $\vec\xi_\perp=\xi^+=0$.
The approach we shall take has been described in detail in Refs.
3-5 for several models, and consists of a straightforward
evaluation of the matrix elements in Eq. 1,
using a Peierls-Yoccoz momentum eigenstate\cite{6} to describe the nucleon
in its rest frame. While
other approaches exist in the literature\cite{7}, our belief is that
the current method maintains closer contact with the quark model in
question, and is more easily generalized to systems of more than
one nucleon, which we will examine elsewhere.

	Phenomenologically, a problem arises when the quark distributions
are evaluated using the unperturbed, SU(4) symmetric nucleon wavefunction.
As a result of the symmetry, the u and d valence quark distributions have
the same functional form, and consequently $d_v(x)/u_v(x)=1/2$ for all
$x$. Experimentally, it is well known that the ratio decreases as $x$
increases, and is thought to vanish linearly as $x\rightarrow 1$. The
missing ingredient, it seems, is a mechanism for SU(4) breaking.

Fortunately, two mechanisms immediately present themselves. The first
is the idea that the nucleon is surrounded by a cloud of virtual mesons,
mainly pions, and that the pions contribute a flavor asymmetry to both
the valence and sea distributions\cite{7.5}. While this idea is certainly
worthy of study, it is not the subject of the current effort. Instead, we
shall concentrate on the second mechanism for SU(4) breaking, the color
magnetic interaction.

The essential idea here is that the spin-spin
interaction results in a dependence of the quark wavefunctions on the spin
state of the other quarks that reside in the nucleon. Since, in the naive
SU(4) wavefunction of the nucleon, the spin is unevenly distributed among
the different flavor quarks, the spin dependence of the color magnetic
interaction is transmuted into a flavor dependence of the spin averaged
quark wavefunctions. Put into the calculation of quark distributions,
this flavor dependence in the wavefunctions turns into a flavor dependence
of the quark momentum distributions.

	In this paper, we calculate the color magnetic corrections to
the valence distributions of the nucleon using the Los Alamos Model
Potential(LAMP). In the next section, we briefly describe the LAMP,
and calculate its quark distributions in the absence of color magnetic
effects. We then begin the calculation of the color magnetic corrections,
separating out contributions to the valence distributions involving
two and three body effects, respectively. We then describe
the contributions of the gauge correction to the valence distributions,
and how they may be partially resummed to obtain improved distributions.
Finally, we shall evolve the full expressions for the corrected $u$
and $d$ distributions to high $Q^2$, where they will be compared with
experiment, and discuss the results.

\section{Two and Three body Corrections}

	In order to set the stage for the forthcoming ${\cal O}(\alpha_s)$
calculations, we begin by calculating the unperturbed quark distributions
for the LAMP model\cite{8}. The model consists of three massless quarks, bound
together in a linear scalar potential,
\begin{equation}
V(r)=V_0(r-r_0),
\end{equation}
with parameters $V_0$= 0.9 GeV/fm and $r_0$=0.57 fm. Wavefunctions
and single particle energies are obtained by solving the Dirac equation
for this potential. As is usual for quark models, the potential parameters
are chosen to generate the average of the nucleon and delta masses, and
the color magnetic interaction is invoked to generate the nucleon-delta
splitting.

	In Ref. 5, the matrix elements in Eq. 1 are evaluated,
assuming that the nucleon is described by a Peierls-Yoccoz momentum
eigenstate,
\begin{equation}
\vert N,P=\vec 0\rangle =\lambda \int d^3a \vert N, \vec R_{cm}=\vec a\rangle,
\end{equation}
where $\lambda$ is a constant required to covariantly normalize the state,
and\\ $\vert N \vec R_{cm}=\vec a\rangle$ denotes the unprojected state
with center fixed at $\vec R_{cm}=\vec a$. If one assumes that the time
dependence
of the quark field operators is well approximated by the the single particle
eigenvalue of the Dirac equation, the expression for the valence quark
distribution is given by
\begin{eqnarray}
q_V^i(x)&=& {MN_i\over \pi V}\Big{[}\big{[}\int_{\vert k_-\vert}^\infty
\, dk\, G(k)\,\big{(} t_0^2(k)+t_1^2(k)+{2k_-\over k}t_0(k)t_1(k)\big{)}\big{]}
\nonumber\\
& &\qquad\qquad\qquad\qquad +\big{[} k_- \rightarrow k_+\big{]}\Big{]},
\end{eqnarray}
where
\begin{eqnarray}
G(k)&=&\int r\, dr\, \sin kr\, \Delta^2(r) EB(r), \nonumber\\
V&=&\int r^2\,dr \Delta^3(r)EB(r),\nonumber\\
t_0(k)&=&\int r^2\,dr j_0(kr)u(r),\nonumber\\
t_1(k)&=&\int r^2\,dr j_1(kr)v(r),\nonumber\\
\Delta(r)&=&\int d^3z \psi_0^\dagger(\vec z-\vec r)\psi_0(\vec z),\nonumber\\
& &
\end{eqnarray}
with $\psi_0(\vec r)$ the ground state, single quark wavefunction,
with upper and lower components $u(r)$ and $i\vec\sigma\cdot\vec r v(r)/r$,
$k_{\pm}=\omega\pm Mx$, with $\omega$ the single particle energy, and
$EB(r)=\langle EB, \vec R_{cm}=r\vert EB,\vec R_{cm}=\vec 0\rangle$ is
the matrix element between two ``empty bags'' separated by a distance $r$,
which accounts for the dynamics of the confining forces.
In the LAMP picture, the dynamics of the confining fields are not specified,
so we have no guidance on the proper choice for this function\cite{4a}.
For the
remainder of this paper, we assume $EB(r)\approx const$.
The resulting distribution is plotted, along with those of the MIT\cite{3}
and soliton\cite{5} bag
models in figure 1. The LAMP distribution lies between the MIT bag, where
the valence quarks carry all of the momentum at the bag scale, and the
soliton bag, where roughly a quarter of the proton's momentum is carried
by the confining degrees of freedom. We expect, therefore, that the quark
model scale, $\mu_0^2$, from which we need to evolve the LAMP distribution
in order to compare with experiment, will lie somewhat below the .6 GeV$^2$
found in reference 5 for the soliton bag. If we vary our assumption
of a constant $EB(r)$, it is easy to generate distributions that interpolate
continuously between the curve shown in figure 1 and curves resembling the
soliton bag distribution. Correspondingly, the scale $\mu_0^2$ will increase
to a value in the neighborhood of .6 GeV$^2$.

	Before calculating the color magnetic corrections to
Eq. 4, we must engage in a brief digression on the questions of
renormalization prescriptions and gauge choices. As it stands, Eq. 4
represents the results of a QCD inspired model of the structure of the
nucleon. If the model were, in fact, a solution of the QCD Hamiltonian
in some calculational scheme, the choice of gauge and renormalization
scheme would be specified, and the quark model wavefunctions would be
unambiguously defined. Unfortunately, there is no known scheme for
generating the LAMP(or any other) model directly from QCD, and consequently
the gauge and renormalization schemes, if any, that produce the wavefunctions
of reference 9 are unspecified. In order to completely specify the model, it
is necessary to postulate
that the model wavefunctions are calculated in a {\it particular} gauge,
with a {\it particular} choice of renormalization scheme. As these choices are
necessarily ad hoc, it follows that different choices will, in general,
result in different models, and necessarily different predictions for
physical observables. For the purposes of
this paper, we shall choose the $\overline{MS}$ renormalization scheme, and
$A^0=0$ gauge.

	The strategy we shall employ for evaluating the color magnetic
corrections to the valence quark distributions is a straightforward
application of perturbation theory, augmented by a closure approximation
whose scale is the mean energy of the gluon in the nucleon. The
interaction Hamiltonian is given by
\begin{equation}
H_I=g_s\int d^3r \vec A_a(\vec r,t)\vec J^a(\vec r,t),
\end{equation}
where $g_s$ is the strong coupling constant, $\vec A_a(\vec r,t)$ is the
gauge field, and $\vec J^a(\vec r,t)= \bar\psi(\vec r,t){\lambda^a\over 2}
\vec\gamma\psi(\vec r,t)$ is the colored quark current operator, with
$\lambda^a\over 2$ an SU(3) generator, and $\vec\gamma$ a Dirac matrix.
To leading order in $g_s$, the nucleon wavefunction is given by
\begin{equation}
\vert N, \vec P=\vec 0\rangle =Z^{-{1\over 2}}\big{[}\vert N_0,\vec P=0\rangle
+{1\over E_0-H_0}H_I\vert N_0,\vec P=0\rangle\big{]},
\end{equation}
where $Z$ is a wavefunction renormalization constant, and $E_0$ is the ground
state energy before correction. The energy shift is given by
\begin{equation}
\delta E={1\over 2E_0V_{\infty}}\langle N_0,\vec P=\vec 0\vert H_I
{1\over E_0-H_0}H_I\vert N_0, \vec P=\vec 0\rangle,
\end{equation}
with $V_{\infty}$ the volume of space, and the renormalization constant
$Z$ is
\begin{equation}
Z=1+{1\over 2E_0V_\infty}\langle N_0,\vec P=\vec 0\vert H_I{1\over
(E_0-H_0)^2}H_I\vert N_0, \vec P=\vec 0\rangle
\end{equation}

	Inserting a complete set of gluon states, and a complete set of
colored, 3 quark states, and using the fact that the unperturbed Hamiltonian
from Eq. 2 is independent of color, the energy denominators may be seen to
be given by
\begin{equation}
H_0-E_0=E_R(\vec k)+\omega(\vec k)-E_0,
\end{equation}
where $\vec k$ is the gluon momentum, $\omega(\vec k)$ its energy,
and $E_R(\vec k)$ is the energy of an excited 3 quark state $R$ with
momentum $-\vec k$.  In order to parallel the assumptions regarding
the gluon propagator in reference 9,
and to facilitate evaluation of the numerical integrations to come,
we make the ansatz that $\omega(\vec k)=\mu e^{k^2/2\mu^2}$, where
$\mu$ is an effective gluon mass taken to be 400 MeV. The energy
denominator is then given by
\begin{equation}
H_0-E_0=\mu e^{k^2/2\mu^2}(1+e^{-k^2/\mu^2}(E_R(\vec k)-E_0)/\mu).
\end{equation}
A reasonable expectation is that the OGE process couples dominantly
to the ground state, and that the relevant gluon momenta are small
(on the order of $1/R_p$), so that the recoil energy of the massive nucleon
is small, and the second term in parentheses
can be neglected. With this assumption,
the energy denominator becomes independent of the intermediate three
quark excited state, and closure may be invoked to perform the sum
over excited states. The results of equations 7-9 keep the same
form , except that the factors $E_0-H_0$ are replaced by the
constant $\omega(\vec k)$.

	Setting ${\bf P}_\pm= 1$, the ${\cal O}(\alpha_s)$ corrections
to eq. 1 may be obtained using  eq. 7. The result is
\begin{eqnarray}
\delta q(x) &=&-Z^{-1}{1\over\mu^2}\int d\xi^-\, e^{-iMx\xi^-/\sqrt{2}}
\int d^3r d^3r^\prime\, G(\vec r-\vec r^\prime)\nonumber\\
& &\times\langle N_0,\vec P=\vec 0
\vert \vec J_a(\vec r)\bar\psi(\xi^-)\gamma^+\psi(0)\vec J_a(\vec r^\prime)
\vert N_0,\vec P=\vec 0\rangle\vert_{LC}\nonumber\\
& &+(Z^{-1}-1)q^0(x),
\end{eqnarray}
where $J_a(\vec r) =\bar\psi(\vec r){\lambda_a\over 2}\psi(\vec r)$,
$G(\vec r)=Ae^{-\mu^2r^2/4}$ is the effective gluon propagator in
the closure approximation, with $A\propto\alpha_s={g_s^2\over 4\pi}$ an
overall strength parameter chosen
to reproduce the nucleon-delta mass splitting, and $q^0(x)$ is the
unperturbed quark distribution of eq. 4.

\subsection{Two-Body Correction}

	In general, eq. 12 gives rise to three types of terms which
may be classified by the number of valence quarks acted on by the three
currents appearing in the matrix element. The ``one-body'' terms, in which
all three currents act on the same quark, represent self interactions and
are traditionally ignored in quark model calculations. Here, these
corrections will tend to mimic the effects of QCD evolution, and may be
absorbed into a redefinition of
the scale $\mu_0^2$ where the evolution begins. Since a precise
determination of this parameter is not the purpose of the present
effort, we shall honor tradition and neglect the one-body terms.

Next, there are the ``two-body'' terms, which arise when a gluon
is in the process of being exchanged between two valence quarks when
one of these quarks is struck by an external probe, such as a photon.
Finally, there are three-body terms, in which the two spectator quarks
exchange a gluon while the struck quark interacts with the external probe.
These terms cannot be absorbed into a redefinition of $\mu_0$, since as we
shall see, they are dependent on the isospin of the struck quark.

	Using fermion anticommutation relations, the two and three-
body corrections to the valence quark distributions can be extracted in
a straightforward manner. The result for the two-body correction is
\begin{eqnarray}
\delta q^{2b}_{\alpha}(x)&=&{\lambda^2\over \mu^2}\sum_{\beta\neq\alpha}\langle
{\lambda_a^\alpha\lambda^{a\beta}\over 4}\rangle\int{d^3k_1d^3k_2\over
 (2\pi)^6}
(2\pi)\delta({Mx-\omega-\hat z\cdot \vec k_1\over \sqrt{2}})\nonumber\\
& &\times \Big{[} \vec Z_\beta(\vec k_1,\vec k_2)\cdot\bar\psi_{0\alpha}(\vec
k_2)
\vec\gamma\gamma^0\gamma^+\psi_{0\alpha}(\vec k_1) +
(\vec k_1 \leftrightarrow \vec k_2,\gamma^+\leftrightarrow\vec\gamma)\Big{]},
\nonumber\\
& &
\end{eqnarray}
where $\alpha$ denotes the struck quark, $\beta$ the spectator with whom the
gluon is exchanged, $\psi_0(\vec k)$ is the quark
momentum space wavefunction, the brackets denote an average over the color
wavefunction of the proton, and
\begin{equation}
\vec Z_\beta(\vec k_1,\vec k_2)=\int d^3r_1d^3r_2\, e^{i(\vec k_1\cdot\vec r_1
+\vec k_2\cdot\vec r_2)} \vec Z_\beta(\vec r_1,\vec r_2)\Delta
(\vec r_1-\vec r_2) EB(\vec r_1-\vec r_2),
\end{equation}
with
\begin{equation}
\vec Z_\beta(\vec r_1,\vec r_2)=\int d^3x\, G(\vec x)\bar\psi_{0\beta}
(\vec x-\vec r_1)\vec\gamma\psi_{0\beta}(\vec x-\vec r_2).
\end{equation}
In order to extract the physically relevant $u$ and $d$ quark distributions,
we must average over the spins in eqs. 13-15.To do the average, we must
separate the spin dependence of the quark wavefunctions appearing in eq.
13. For s-wave quark wavefunctions, we get
\begin{eqnarray}
\bar\psi_{0\alpha}(\vec k_1)\vec\gamma\gamma^0\gamma^+\psi_{0\alpha}(\vec k_2)
&=&\vec k_1 C_{01}(k_2,k_1)-\vec k_2C_{01}(k_1,k_2)\nonumber\\
&   &+i((\vec k_2\times \vec\sigma_\alpha)C_{01}(k_1,k_2)+
(\vec k_1\times \vec\sigma_\alpha)C_{01}(k_2,k_1))\nonumber\\
&   &+\hat z(C_{00}(k_1,k_2)-(\vec k_1\cdot\vec k_2+i\vec k_2\cdot(\vec k_1
\times \vec\sigma_\alpha))C_{11}(k_1,k_2))\nonumber\\
&&+i\hat z\times( \vec\sigma_\alpha (C_{00}(k_1,k_2)+\vec k_1\cdot\vec k_2
C_{11}(k_1,k_2))\nonumber\\
&   &-(\vec k_2\vec k_1\cdot \vec\sigma_\alpha+\vec k_1\vec k_2\cdot\vec
\sigma_\alpha+i(\vec k_1\times\vec k_2))C_{11}(k_1,k_2))),
\end{eqnarray}
where
\begin{equation}
C_{ij}(k_1,k_2)={4\pi\over\sqrt{2}}{t_i(k_1)t_j(k_2)\over k_1^ik_2^j},
\end{equation}
and $(i,j)\in (0,1)$ as defined in eq. 5.
The function $\vec Z_\beta(\vec k_1,\vec k_2)$ can be decomposed
similarly,
\begin{eqnarray}
\vec Z_\beta(\vec k_1,\vec k_2)& =&\vec k_1 Z^1_\beta (\vec k_1,\vec k_2)
+\vec k_2 Z^2_\beta(\vec k_1,\vec k_2)\nonumber\\
& &+i((\vec k_1\times \vec\sigma_\beta)
Z^3_\beta(\vec k_1,\vec k_2)+(\vec k_2\times\vec\sigma_\beta)
 Z^4_\beta(\vec k_1,\vec k_2)),
\end{eqnarray}
where the $Z^i_\beta(\vec k_1,\vec k_2)$ are scalar functions of $\vec k_1$
and $\vec k_2$, evaluated semi-analytically using eq. 15 and the six gaussian
fit to the quark wavefunctions of reference 9.

	Once the integrations are performed, there are only three vectors left,
$\hat z$,$\vec\sigma_\alpha$ and $\vec\sigma_\beta$, and the two-body
correction takes the form
\begin{equation}
\delta q^{2b}_{\alpha}(x)= f^{2b}(x)+\sum_{\beta\neq\alpha}(
\vec\sigma_\alpha^i\vec\sigma_\beta^jT^{2b}_{ij}(x),
\end{equation}
where $T^{2b}_{ij}(x)$ is a tensor defined in terms of the
quark wavefunctions.

	Using rotational arguments, it is easy to demonstrate that only the
trace of $T$ will contribute to the spin independent quark distribution. Hence,
the functions $f^{2b}(x)$ and $T_{ii}(x)$ may be written as
\begin{eqnarray}
f^{2b}(x)&=&-{16 Z^{-1}\lambda^2\over 3\mu^2}\int {d^3k_1d^3k_2\over (2\pi)^5}
\delta(Mx-\omega-\hat z\cdot\vec k_1)\nonumber\\
& &\times\Big{[}Z^2(\vec k_1,\vec k_2)\Big{(}k_1^2C_{10}(k_1,k_2)+\vec
k_1\cdot\vec k_2 C_{01}(k_1,k_2)\nonumber\\
& &\qquad\qquad\qquad
+\hat z\cdot\vec k_1(C_{00}(k_1,k_2)+\vec k_1\cdot\vec k_2
C_{11}(k_1,k_2))\Big{)}\nonumber\\
& &+Z^1(\vec k_1,\vec k_2)\Big{(}\vec k_1\cdot\vec k_2C_{10}(k_1,k_2)+k_2^2
C_{01}(k_1,k_2)\nonumber\\
& &\qquad\qquad\qquad
+\hat z\cdot\vec k_1({\vec k_1\cdot\vec k_2\over k_1^2}
C_{00}(k_1,k_2)+k_2^2
C_{11}(k_1,k_2))\Big{)}\Big{]},\nonumber\\
\end{eqnarray}
and,\\
\begin{eqnarray}
T^{2b}_{ii}(x)&=&-{32Z^{-1}\lambda^2\over 3\mu^2}
\int {d^3k_1d^3k_2\over (2\pi)^5}
\delta(Mx-\omega-\hat z\cdot\vec k_1)\nonumber\\
& &\times\Big{[}Z^4(\vec k_1,\vec k_2)\Big{(}
k_1^2C_{10}(k_1,k_2)-\vec k_1\cdot\vec k_2C_{01}(k_1,k_2)
\nonumber\\
& &+\hat z\cdot\vec k_1(C_{00}(k_1,k_2)
-\vec k_1\cdot\vec k_2 C_{11}(k_1,k_2)\Big{)}\nonumber\\
& &+Z^3(\vec k_1,\vec k_2)\Big{(}\vec k_1\cdot\vec k_2 C_{10}(k_1,k_2)-
k_2^2C_{01}(k_1,k_2)
\nonumber\\& &+\hat z\cdot \vec k_2(C_{00}(k_1,k_2)-\vec k_1\cdot\vec
k_2C_{11}(k_1,k_2)\Big{)}\Big{]},\nonumber\\
\end{eqnarray}
	Assuming that the unperturbed proton wavefunction is SU(4)
symmetric, the spin sums are given by
\begin{equation}
\sum_{\alpha\neq\beta} \vec\sigma_\alpha\cdot\vec\sigma_\beta
{(1\pm\tau_{3\alpha})\over 2}=-3\pm 1,
\end{equation}
so that the two-body corrections to the physical quark distributions are
given by
\begin{eqnarray}
\delta u^{2b}(x)&=&2f^{2b}(x)-2T^{2b}_{ii}(x)\\
\delta d^{2b}(x)&=&f^{2b}(x)-4T^{2b}_{ii}(x).
\end{eqnarray}
	An interesting feature of this result is that the $d$ quark
distribution is twice as sensitive to color magnetic effects
than the $u$ distribution. On physical grounds, this is understandable,
since the two spectator $u$ quarks are necessarily in a spin 1 state in
the SU(4) wavefunction, while the spectator $ud$ pair is most likely
to be found in a spin 0 state. Numerically, the color electric correction,
$f^{2b}(x)$, is quite small and is neglected in the following.

The correction to the antiquark distribution is obtained from
eq. 13 by replacing $\omega\leftrightarrow -\omega$ and including an
overall minus sign.

\subsection{ Three Body Correction}

	After separating out the one and two-body terms from equation 12,
only the three-body term remains. Physically, this correction accounts
for the modified structure of the spectator diquark, and is expressed
as a modification of the recoil function $G(k)$ appearing in equation
4. In terms of the quark wavefunctions the three body correction is given
by
\begin{eqnarray}
\delta q^{3b}_V(x)&=&{Z^{-1}\lambda^2\over \mu^2}
\sum_{\alpha\neq\beta\neq\epsilon}\langle
{\lambda_a^\epsilon\lambda_a^\beta\over 4}\rangle\int {d^3k\over (2\pi)^3}
(2\pi)\delta({Mx-\omega-\hat z\cdot \vec k\over \sqrt{2}})\nonumber\\
& &\times\bar\psi_{0\alpha }(\vec k)\gamma^+\psi_{0\alpha }(\vec k)
F_{\beta\epsilon}(-\vec k),
\end{eqnarray}
where the function $\tilde F_{\beta\epsilon}(\vec k)$ is given by
\begin{equation}
\tilde F_{\beta\epsilon}(\vec k)=\int d^3y e^{i\vec k\cdot \vec y}
F_{\beta\epsilon}(\vec y)EB(y),
\end{equation}
with
\begin{equation}
F_{\beta\epsilon}(\vec z)=\int\int d^3r_1d^3r_2 G(\vec r_1-\vec r_2)
\bar\psi_{0\beta}(\vec r_1)\vec\gamma\psi_{0\beta}(\vec r_1-\vec z)\cdot
\bar\psi_{0\epsilon}(\vec r_2)\vec\gamma\psi_{0\epsilon}
(\vec r_2-\vec z).
\end{equation}

	Just like the two-body correction, $F_{\beta\epsilon}(\vec z)$
can be decomposed into a scalar function, $f^{3b}(z)$ and the trace of a
tensor $T^{3b}_{ij}(\vec z)$,
\begin{equation}
  F_{\beta\epsilon}(\vec z)=f^{3b}(z)+{\vec \sigma_\beta\cdot\sigma_\epsilon
\over 3}T^{3b}_{ii}(z),
\end{equation}
where
\begin{eqnarray}
f^{3b}(z)&=&-2\int\int d^3r_1d^3r_2 G(\vec r_1-\vec r_2)
\Big{[}v(r_1)v(r_2)u(\vec r_1-\vec z)u(\vec r_2-\vec z)\hat r_1\cdot\hat
r_2\nonumber\\
& &\qquad\qquad -v(r_1)v(\vec r_2 -\vec z)u(r_2)v(\vec r_2-\vec z)
\hat r_1\cdot {(\vec r_2 -\vec z)\over \vert \vec r_2 -\vec z\vert }
\Big{]},
\end{eqnarray}
and,
\begin{eqnarray}
T^{3b}_{ii}(z)&=&4\int\int d^3r_1d^3r_2 G(\vec r_1-\vec r_2)
\Big{[}v(r_1)v(r_2)u(\vec r_1-\vec z)u(\vec r_2-\vec z)\hat r_1\cdot\hat
r_2
\nonumber\\ \qquad\qquad & &+v(r_1)v(\vec r_2 -\vec z)u(r_1)
v(\vec r_2-\vec z)
\hat r_1\cdot {(\vec r_2-\vec z)\over\vert \vec r_2 -\vec z\vert}
\Big{]},
\end{eqnarray}
where $u$ and $v$ are, again, the upper and lower components of the single
quark wavefunctions as in eq. 5.

	The spin sum, including the projection onto quark isospin, is given
by
\begin{equation}
\sum_{\alpha\neq\beta\neq\epsilon} {(1\pm \tau_{3\alpha})\over 2}\vec
\sigma_\beta\cdot\sigma_\epsilon ={ -3\mp 5\over 2}
\end{equation}
so that the three body correction is given by
\begin{eqnarray}
\delta u^{3b}(x)&=&2\delta q_E^{3b}(x)-4\delta q_M^{3b}(x)
\nonumber\\
\delta d^{3b}(x)&=&\delta q_E^{3b}(x)+\delta q_M^{3b}(x),
\end{eqnarray}
with
\begin{eqnarray}
\delta q_E^{3b}(x)&=&{Z^{-1}\lambda^2\over \mu^2}
\sum_{\alpha\neq\beta\neq\epsilon}\langle
{\lambda_a^\epsilon\lambda_a^\beta\over 4}\rangle\int {d^3k\over (2\pi)^3}
(2\pi)\delta({Mx-\omega-\hat z\cdot \vec k\over \sqrt{2}})\nonumber\\
& &\qquad\times\bar\psi_{0\alpha }(\vec k)\gamma^+\bar\psi_{0\alpha }(\vec k)
\tilde f^{3b}(-\vec k),
\end{eqnarray}
and
\begin{eqnarray}
\delta q_M^{3b}(x)&=&{Z^{-1}\lambda^2\over \mu^2}
\sum_{\alpha\neq\beta\neq\epsilon}\langle
{\lambda_a^\epsilon\lambda_a^\beta\over 4}\rangle\int {d^3k\over (2\pi)^3}
(2\pi)\delta({Mx-\omega-\hat z\cdot \vec k\over \sqrt{2}})\nonumber\\
& &\qquad\times\bar\psi_{0\alpha }(\vec k)\gamma^+\bar\psi_{0\alpha }(\vec k)
\tilde T^{3b}_{ii}(-\vec k),
\end{eqnarray}
and where $\tilde f^{3b}(\vec k)$, $\tilde T^{3b}_{ii}(\vec k)$ are the
Fourier transforms
of $EB(z)f^{3b}(\vec z)$ and $EB(z)T^{3b}_{ii}(\vec z)$, respectively.
Again the relative weights appearing in the eq. 32 reflect the fact that
a $ud$ spectator pair is most likely to be in a spin zero state.

\section{Gauge Correction}
	As discussed in the last section, a gauge and renormalization
scheme must be assumed in order for quark model wavefunctions to be
meaningfully defined. Correspondingly, the effect of the path-ordered
exponential of eq. 1 must be included in order to obtain quark distributions
that are gauge invariant. Expanding the path ordered exponential $P_\pm(\xi)$
to leading order in $g_s$, we obtain
\begin{eqnarray}
\delta q_{gauge}(x)&=&
{i\mu}\int d\xi^{-} \int_0^{\xi^-}d\eta^-
\langle N_0,\vec P=\vec 0\vert \bar\psi(\xi^-)A^+(\eta^-)\psi(0)
\nonumber\\& &\qquad\qquad\qquad\qquad\times\int d^3r
\vec J(\vec r)\cdot \vec A(\vec r)\vert N_0,\vec P=\vec 0\rangle.
\end{eqnarray}
	Neglecting self interactions, this expression may be evaluated
in the same fashion as the two and three body corrections. The result,
after some manipulation, is given by
\begin{eqnarray}
\delta q_{gauge}(x)&=& {16\lambda^2\sqrt{2}\over\mu M}{d\over dx}
\sum_{\alpha\neq\beta}
\int_0^1 dy\int {d^3Pd^3q\over (2\pi)^5}\delta(Mx-\omega-\hat z\cdot q)
\nonumber\\
& &\times\bar\psi_\alpha(y\vec P-\vec q)\gamma^+
\psi_\alpha((1-y)\vec P+\vec q)\hat z\cdot \vec Z_\beta(y\vec P-\vec
q,(1-y)\vec P+\vec q).\nonumber\\
& &
\end{eqnarray}
	The spin structure of the gauge correction is identical to that of
the two-body correction, so we may immediately write
\begin{eqnarray}
\delta u_{gauge}(x)&=&{d\over dx}(2f^{gauge}(x) + 2T^{gauge}_{ii}(x))
\nonumber\\
\delta d_{gauge}(x)&=&{d\over dx}(f^{gauge}(x) + 4T^{gauge}_{ii}(x)),
\end{eqnarray}
where
\begin{eqnarray}
f^{gauge}(x)&=&{16\lambda^2\over\mu M}\int_0^1 dy\int {d^3Pd^3q\over
(2\pi)^5}\delta(Mx-\omega-\hat z\cdot q)\nonumber\\
& &\times\Big{[}Z^1(\vec k_2,\vec k_1)\hat z\cdot\vec k_2(\hat z\cdot\vec k_2
C_{10}(k_1,k_2)+\hat z\cdot\vec k_1 C_{01}(k_1,k_2)\nonumber\\
& &\qquad\qquad\qquad+(C_{00}(k_1,k_2)+\vec k_1\cdot\vec k_2)C_{11}(k_1,k_2))
\nonumber\\
& &+Z^2(\vec k_2,\vec k_1)\hat z\cdot\vec k_1(\hat z\cdot\vec k_2
C_{10}(k_1,k_2)+\hat z\cdot\vec k_1 C_{01}(k_1,k_2)\nonumber\\
& &\qquad\qquad\qquad+(C_{00}(k_1,k_2)
+\vec k_1\cdot\vec k_2C_{11}(k_1,k_2)))\Big{]},
\nonumber\\
\end{eqnarray}
\begin{eqnarray}
T^{gauge}_{ii}(x)&=&-{16\lambda^2\over 3\mu M}\int_0^1 dy\int {d^3Pd^3q
\over (2\pi)^5}\delta(Mx-\omega-\hat z\cdot\vec q)\nonumber\\
& &\times\Big{[}Z^3(\vec k_2,\vec k_1)
((\vec k_1\cdot\vec k_2-\hat z\cdot\vec k_1\hat z\cdot\vec k_2)C_{01}(k_1,k_2)
\nonumber\\
& &
-(k_2^2-(\hat z\cdot \vec k_2)^2)C_{10}(k_1,k_2)+(\hat z\cdot\vec k_2
\vec k_1\cdot\vec k_2 -k_2^2\hat z\cdot\vec k_1)C_{11}(k_1,k_2))\nonumber\\
& &-Z^4(k_2,k_1)((\vec k_1\cdot\vec k_2-\hat z\cdot\vec k_1\hat z\cdot\vec k_2)
C_{10}(k_1,k_2)
\nonumber\\
& &-(k_1^2-(\hat z\cdot \vec k_1)^2)C_{01}(k_1,k_2)
-(k_1^2\hat z\cdot\vec k_1-\vec k_1\cdot\vec k_2\hat z\cdot\vec k_1)
C_{11}(k_1,k_2))\Big{]}\nonumber\\
& &
\end{eqnarray}
with $\vec k_1=(1-y)\vec P+\vec q$ and $\vec k_2=y\vec P -\vec q$.

	The gauge factor in eq. 1 removes the unphysical,
gauge dependent phase in the quark wavefunction, which produces a mismatch
between the momentum fraction $x$ calculated using simple Fourier transforms
and the physical momentum carried by the struck quark. This shift in $x$ is
manifested by the derivative with respect to $x$ appearing in the gauge
correction. This pattern persists in higher orders, as the path ordered
exponential is expanded in powers of $g_s{d\over dx}$, and suggests that
the gauge correction may be partially resummed to obtain a quark distribution
of the form
\begin{equation}
q^i(x)=(1+{ds_i(x)\over dx})q^i_0(x+s_i(x)),
\end{equation}
where $q^i_0(x)$ is the zeroeth order quark distribution for quarks of flavor
$i$. To leading order in $\alpha_s$, the shift function $s_i(x)$ is given
by
\begin{equation}
s_i(x)={(a_i f^{gauge}(x)+b_i T^{gauge}_{ii}(x))\over q^i_0(x)},
\end{equation}
with $a_i$ and $b_i$ the isospin dependent coefficients appearing in eq.
37. Since both $\alpha_s$ and the derivatives of quark distributions can
be large in the context of quark models, the effects of the resummation
can be important, particularly for the $d$ distribution.

\section{Results}

	The expressions for the gauge, two and three-body corrections to
the valence distributions of the proton were evaluated using a Gaussian
quadrature scheme to perform the multiple integrations appearing in eqs.
20,21,33,34,38, and 39. For the gauge and two-body corrections, the function
$\vec Z_\beta(\vec k_1,\vec k_2)$ was evaluated semi-analytically by making
use of the 6 gaussian fit to the LAMP coordinate space wavefunctions
described in reference 9. After accounting for the rotational symmetries
of the problem, there were 4,5, and 6 integrations to be performed numerically
for the two-body, gauge and three-body corrections, respectively. Treating
all color magnetic effects perturbatively, we obtain the curves shown in
Figs. 2 and 3 for the $d$ and $u$ valence distributions, respectively.
Also shown are the distributions with no color magnetic corrections applied,
and the curves obtained by
either neglecting the gauge correction, or by resumming the gauge correction
in the fashion described at the end of section 3.

	As shown in Figs. 2 and 3, the general effect of CMI is to lower the
momentum carried by the valence quarks. Physically, this makes sense, since
the exchanged gluons will carry $p^+>0$, and consequently the momentum
fraction carried by the struck quark must decrease. As advertised in the
derivation of the two and three-body corrections, the $d$ quark distribution
is more sensitive to CMI than the $u$ distribution, since the $uu$ spectator
pair is more likely to be found in a spin 1 state than the $ud$ pair. Less
encouraging is the fact that the Peierls-Yoccoz projection procedure is
significantly less effective in producing corrected distributions with
good support properties than it is for the unperturbed distributions. In
particular, the two-body correction to the valence distributions has a
non-negligible tail($\approx .01$) in the region $x\approx 1$. This increase
in the relative size of the tail may be understood in terms of a small
admixture of higher mass eigenstates in the Peierl-Yoccoz projected
nucleon wavefunction. Since these states are more massive, the quarks they
contain are not constrained to carry momentum $p^+<M/\sqrt{2}$ in the nucleon
rest frame, and show up as a tail in the region $x\ge 1$. Since there
are many excited states to couple to, the
size of this unwanted component of the wavefunction will tend to grow
when a perturbation, such as the color magnetic interaction,
is introduced\cite{8.9}.
A more serious problem, from a phenomenological point of view,
is the fact that the gauge correction is comparable to the two and
three-body corrections. This suggests that the $d$ quark
distribution will be quite sensitive to the choice of gauge used to
define quark model wavefunctions. In principle, this provides an
additional restriction on the models, but in practice the infinite
variety of gauges may provide a means to reconcile virtually any model
with the data.

	The quark distributions are evolved from $\mu_0^2$ to 15 GeV$^2$
using a finite element procedure to reconstruct the evolved distributions
from their moments. A detailed description of this procedure will be given
elsewhere\cite{9}. The results for the $d$ and $u$ distributions are shown
in Figs. 4 and 5, respectively, for $\mu_0^2$ ranging from .2-.4 GeV$^2$.
Also
shown are BEBC neutrino data at the same $Q^2$\cite{10}. In general,
the data at large $x$ favor a choice of the scale $\mu_0^2$ between
.2 and .3 GeV$^2$, while at smaller x, the data lies below model predictions
even at these small renormalization scales. This situation may reflect
the uncertainty associated with modeling the sea distributions, which are
used to analyze the data in this region, or it may reflect a shortcoming
of the model, namely that too much momentum is carried by the valence
quarks at the quark model scale. In the former case, we note that
parametrizations of parton distributions based on larger data sets\cite{11}
generate larger valence distributions in the region of $x$ in question.
In the latter case, we have already noted that the momentum fraction carried
by the valence quarks may be altered significantly by changing the spatial
dependence of the empty bag matrix element, $EB(z)$.

	In light of these facts, we conclude that color magnetic interactions
provide a natural mechanism for producing the observed differences between
the $u$ and $d$ valence quark distributions of the proton, and that the
LAMP picture provides a reasonable quantitative description of the valence
distributions once color magnetic effects are included.

\centerline{\bf {Acknowledgements}}
	This work was supported in part by the U.S. Department of
Energy, Division of High Energy and Nuclear Physics, ER-23.

\vfill
\eject
\centerline{\bf Figure Captions}
\begin{itemize}
\item{} Figure 1 - Valence quark distributions for the MIT bag, soliton
bag, and LAMP at the quark model scale.
\item{} Figure 2 - Valence distributions for the $d$ quark at the quark model
scale, including color magnetic corrections.
\item{} Figure 3 - Valence distributions for the $u$ quark at the quark model
scale, including color magnetic corrections.
\item{} Figure 4 - Valence distributions for the $d$ quark, evolved to
$Q^2$=15 GeV$^2$. Also shown are data from BEBC.
\item{} Figure 5 - Valence distributions for the $u$ quark, evolved to
$Q^2$=15 GeV$^2$. Also shown are data from BEBC.
\end{itemize}

\begin{thebibliography}{999}
\bibitem{8.9} We note that this problem with perturbation theory
may be avoided if the model is quantized on the light cone, but then the
unperturbed wavefunction becomes difficult to obtain.
\end{thebibliography}
\end{document}